\documentclass{an}
\usepackage{graphicx}
\usepackage{times}
\usepackage{fancyhdr}
\begin{document}

\title{Magnetic fields in jets: ordered or disordered?}

\author{R.A. Laing\inst{1}
\and J.R. Canvin\inst{2} 
\and A.H. Bridle\inst{3}}
\institute{
European Southern
Observatory, Karl-Schwarzschild-Stra\ss e 2, 85748 Garching,
Germany
\and
School of Physics, University of Sydney, A28, Sydney,
NSW 2006, Australia
\and
National Radio Astronomy Observatory, 520 Edgemont Road,
Charlottesville, VA 22903-2475, USA
}

\date{Received; accepted; published online}

\abstract{The question of the degree of order in the magnetic fields of
relativistic jets is important to any understanding of their production. Both
vector-ordered (e.g. helical) and disordered, but anisotropic fields can produce
the high observed degrees of polarization. We outline our models of jets in
FR\,I radio galaxies as decelerating relativistic flows. We then present
theoretical calculations of the synchrotron emission from different field
configurations and compare them with observed emission from FR\,I jets. We show
that large-scale helical fields (with significant poloidal and toroidal
components) are inconsistent with observations. The combination of an ordered
toroidal and disordered poloidal component is consistent with our data, as is an
entirely disordered field. Jets must also contain small, but significant amounts
of radial field.  
}

\correspondence{rlaing@eso.org}

\maketitle

\section{Introduction}

This paper addresses two questions:
\begin{enumerate}
\item Is the magnetic field in extragalactic radio jets vector-ordered or does
it have many reversals?
\item What is its three-dimensional structure?
\end{enumerate}
Observations of polarized synchrotron emission at frequencies where Faraday
rotation is negligible can be used to infer a two-dimensional projection of the
field structure on the plane of the sky (we refer to the {\em apparent magnetic
field direction}, which is the perpendicular to the observed ${\bf E}$-vector
position angle). This emission is an emissivity-weighted integral through the
jet, affected by relativistic aberration if the flow speed is comparable with
$c$ but independent of reversals in the field. Measurement of Faraday
rotation can potentially be used to determine the vector field component along
the line of sight within a jet, but the thermal plasma responsible must be
within the jet volume, rather than in front of it, as usually appears to be the
case (e.g. Laing et al. 2006).

We have developed models of jets in low-power FR\,I (Fanaroff \& Riley 1974)
radio sources as decelerating relativistic flows. By fitting to deep radio
images, we can determine the geometry and the distributions of velocity and
emissivity. Our technique also constrains the three-dimensional structure of the
magnetic field. We cannot determine this structure uniquely, but we can estimate
the ratios of the longitudinal, toroidal and radial components and eliminate
some of the simpler proposed field geometries.

Our fundamental assumption is that jets are intrinsically symmetrical and
relativistic. Aberration then acts differently on radiation from the approaching
and receding jets, so their observed synchrotron images are two-dimensional
projections of the field structure viewed from different directions. From these,
subject to some additional assumptions, we can infer the three-dimensional
structure of the field. One key assumption is axisymmetry: if this holds, we
can also use the symmetry of the transverse brightness and polarization profiles to
distinguish (at least in some cases) between vector-ordered and disordered
fields.

\section{Jet models}

\begin{figure}
\resizebox{\hsize}{!}  {\includegraphics[]{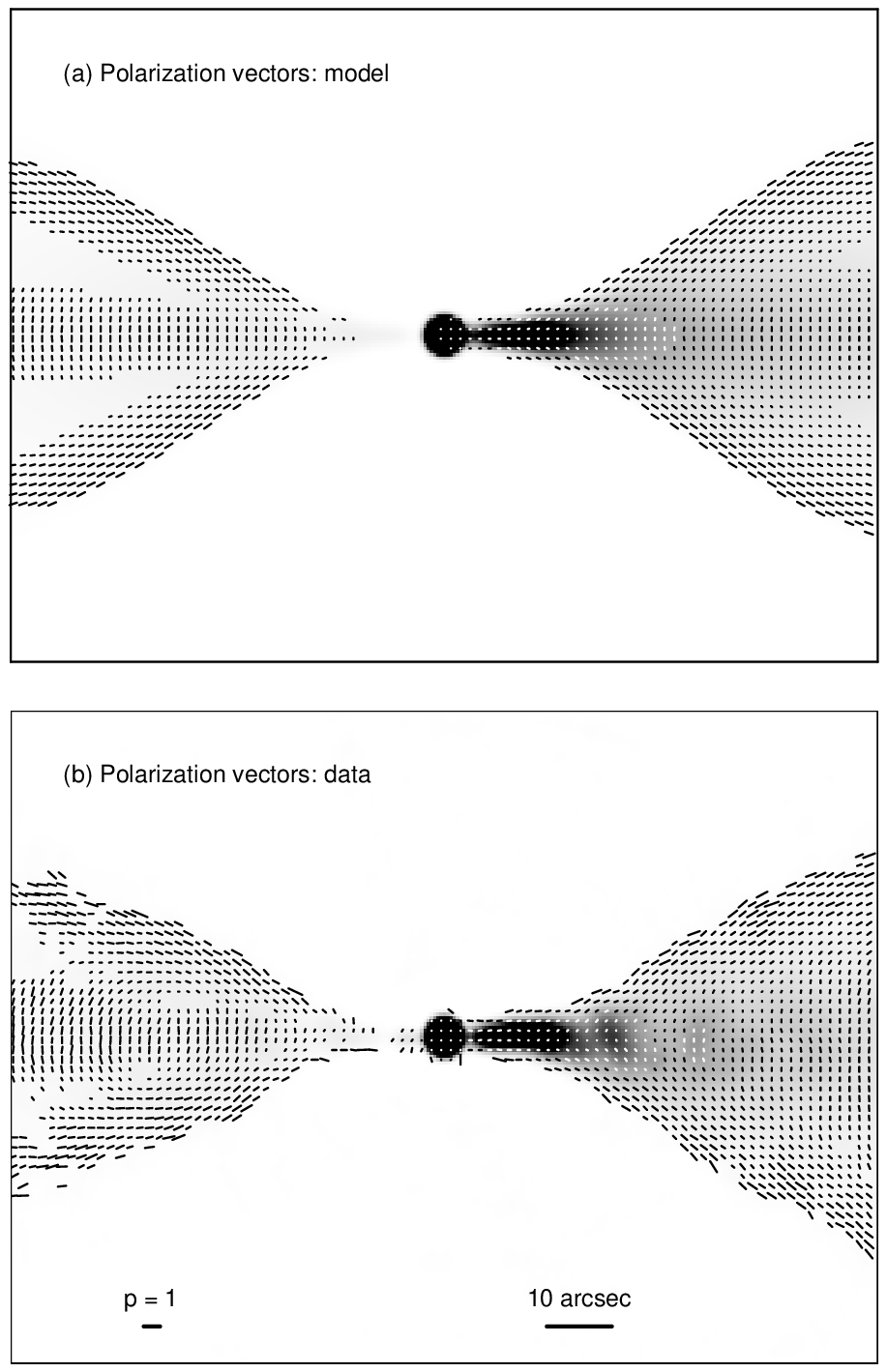}}
\caption{A comparison of VLA data and model for NGC\,315 (Canvin et
al.\ 2005). Vectors with lengths proportional to the degree of polarization, $p$,
and directions along the apparent magnetic field are superimposed on grey-scales
of total intensity. The resolution is 2.35\,arcsec. The polarization and angular
scales are indicated by the labelled bars in the lower panel. (a) model; (b)
data.}
\label{fig:ngc315ivec}
\end{figure}

We model FR\,I jets as intrinsically symmetrical, axisymmetric, relativistic,
stationary flows, in which the magnetic fields are assumed to be disordered, but
anisotropic (see Section~\ref{fieldconf} for a detailed discussion of this
assumption). We adopt simple, parameterized functional forms for the geometry
and the spatial variations of velocity (allowing both deceleration and
transverse gradients), emissivity and field-component ratios. We then optimize
the model parameters by fitting to deep VLA images in $I$, $Q$ and $U$. The
model brightness distributions are derived by integration along the line of
sight, including the effects of anisotropy in the rest-frame emission,
aberration and beaming.  Note that modelling of the linear polarization is
essential to break the degeneracy between angle to the line of sight and
velocity.  Details of the models are given by Laing \& Bridle (2002), Canvin \&
Laing (2004) and Canvin et al.\ (2005).  A comparison between our best-fitting
model and VLA data for the first arcminute of the jets in NGC\,315 is shown in
Fig.~\ref{fig:ngc315ivec} (Canvin et al.\ 2005).

\section{Field configurations}
\label{fieldconf}

The models we have discussed so far fit the observations very well, but make a
specific assumption about the structure of the field: namely that it is
disordered on small scales, but anisotropic. An alternative class of field
structures, proposed on theoretical grounds, has a vector-ordered, quasi-helical
structure. In this section, we consider how to differentiate observationally
between these structures.

\begin{figure}
\resizebox{\hsize}{!}  {\includegraphics[]{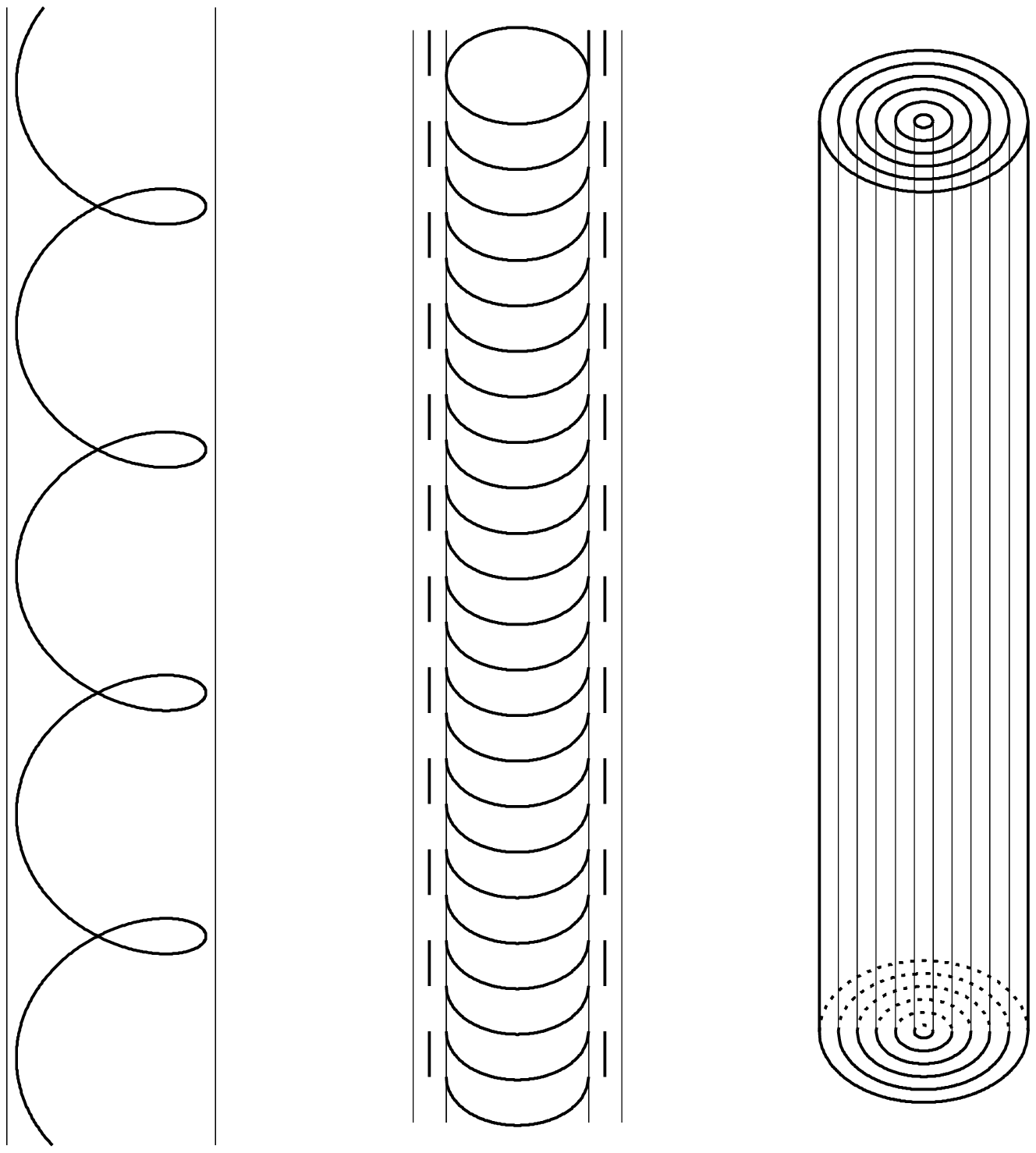}}
\caption{Sketches of the three field configurations discussed in the text. Left:
vector-ordered helix. Middle: perpendicular-field spine and longitudinal-field
shear layer. Right: two-dimensional field sheets wrapped around the jet axis.}
\label{fig:fieldconfigs}
\end{figure}

The generic polarization structure which we seek to model has an apparent
magnetic field which is transverse on the axis of the jet but longitudinal at
its edges (Fig.~\ref{fig:ngc315ivec}). Three field structures which can all
produce this effect are sketched in Fig.~\ref{fig:fieldconfigs}. The first is an
ordered, helical field. For simplicity, we take this to have a constant pitch
angle, but qualitatively similar results are obtained from generically similar,
but more complex configurations. The second has a spine of two-dimensional field
sheets with equal components in directions orthogonal to the jet axis but no
longitudinal component surrounded by a longitudinal-field shear layer (Laing
1980, 1993). The third has two-dimensional field sheets wrapped around the axis,
giving equal toroidal and longitudinal, but no radial component (model B of
Laing 1980).

\begin{figure}
\resizebox{\hsize}{!}  {\includegraphics[]{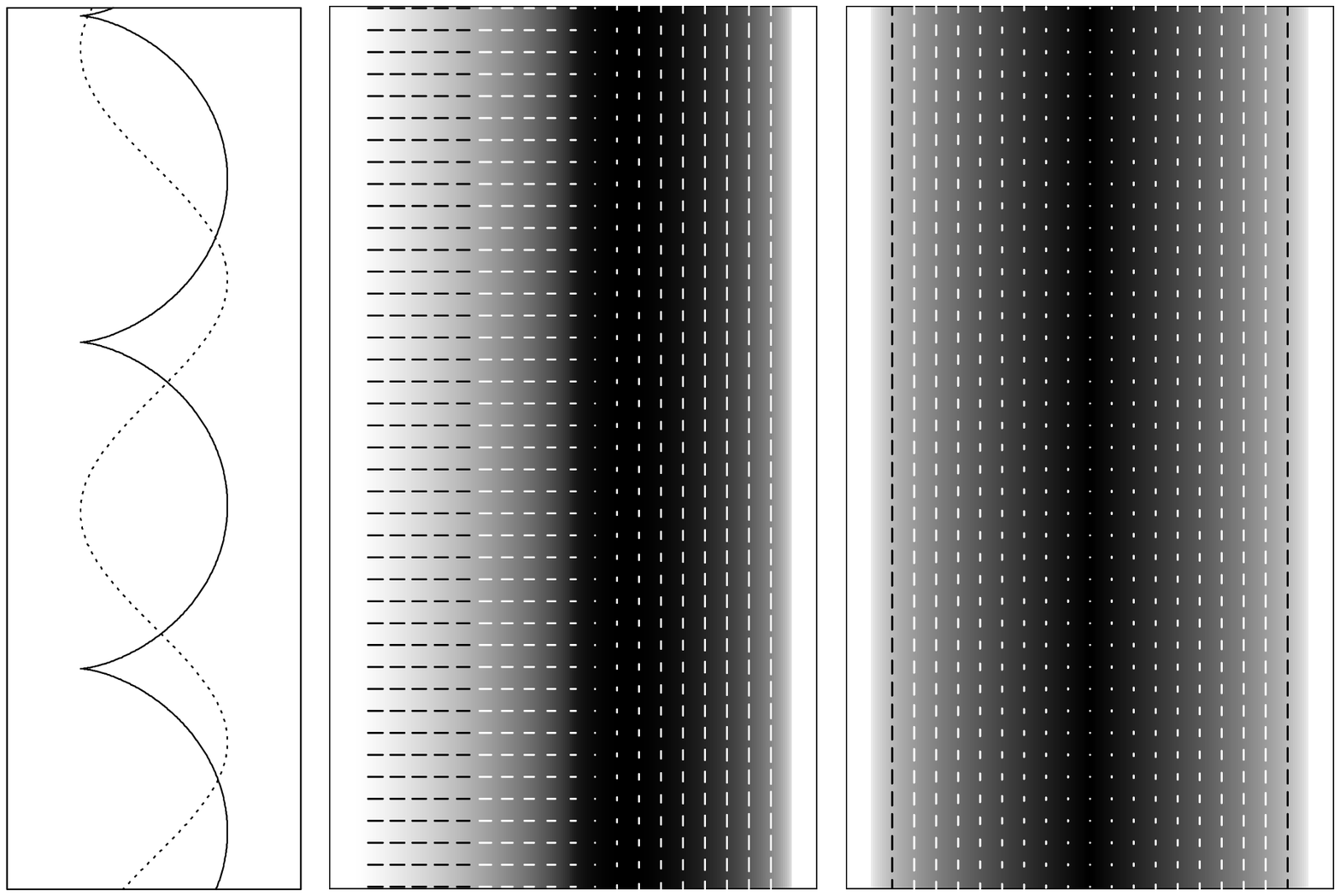}}
\caption{Total intensity and linear polarization from a (non-relativistic) jet
containing a helical magnetic field of pitch angle 45$^\circ$ at angles to the
line of sight of $\theta = 45^\circ$ and 90$^\circ$. Left: sketch showing the
projection of field lines on the plane of the sky. Full line, $\theta =
45^\circ$; dotted line, $\theta = 90^\circ$. Middle and right: grey-scales of total
intensity with superposed vectors whose lengths are proportional to the degree
of polarization and directions along the apparent magnetic field. Middle: $\theta =
45^\circ$; right: $\theta = 90^\circ$.}
\label{fig:helix}
\end{figure}

\begin{figure}
\resizebox{\hsize}{!}  {\includegraphics[]{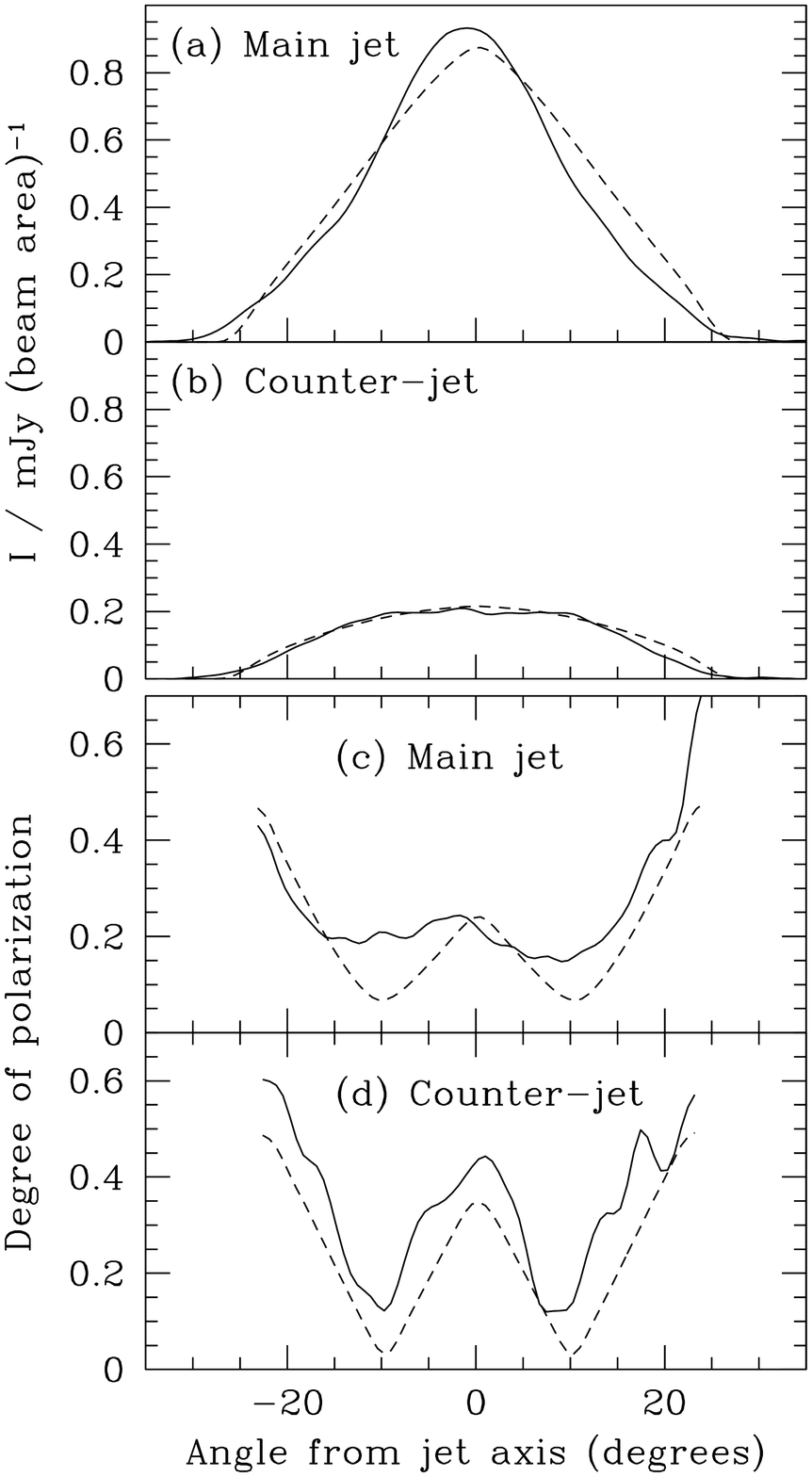}}
\caption{Transverse profiles of total intensity and degree of polarization, $p$,
  for the jets of NGC\,315 (Canvin et al.\ 2005). Full lines: VLA data; dashed
  lines: model. (a) $I$, main jet; (b) $I$, counter-jet; (c) $p$, main jet; (d)
  $p$, counter-jet.}
\label{fig:ngc315transverse}
\end{figure}

Fig.~\ref{fig:helix} illustrates the brightness and polarization structure
produced by a helical field in a jet. In general, helical fields produce
asymmetric transverse brightness and polarization profiles (Laing 1981). The
profiles are symmetrical only if the field is purely toroidal or the jet is at
90$^\circ$ to the line of sight in the rest frame of the emitting material. The
condition for the approaching jet to appear side-on in the rest frame is $\beta
= \cos\theta$\footnote{$\beta = \cos\theta$ is also the condition for maximum
Doppler boost, so there is a selection effect in favour of observing symmetrical
transverse profiles in blazar jets, even if their fields are helical.}.  The
counter-jet can never appear edge-on in its rest frame unless $\beta = 0$ and
$\theta = 90^\circ$. Fig~\ref{fig:ngc315transverse} shows average transverse
profiles of total intensity and degree of polarization for the jets in NGC\,315
(Canvin et al.\ 2005). Our modelling shows that the angle to the line of sight is
$\approx 38^\circ$ and that the field must have a a mixture of longitudinal and
toroidal components.  The profiles are extremely symmetrical, especially in the
counter-jet, so we can rule out an ordered helical field at least in this
source. The toroidal field component could be ordered, for example if the
longitudinal component has many reversals. More generally, a symmetrical
transverse profile requires a symmetrical distribution of field directions in
the rest frame of the jet flow.

\begin{figure}
\resizebox{\hsize}{!}  {\includegraphics[]{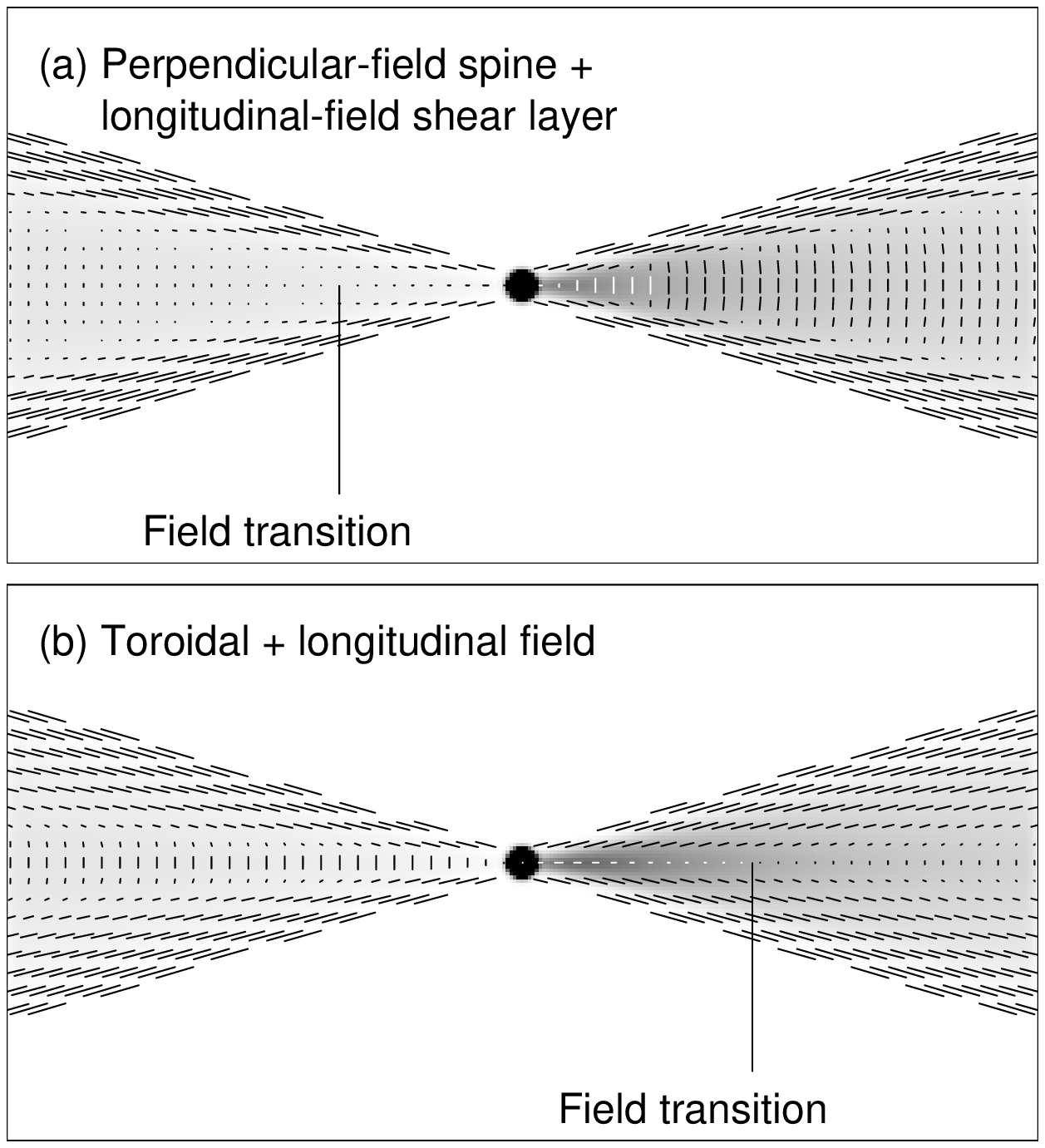}}
\caption{Location of the transition between transverse and longitudinal apparent
magnetic field on the jet axis. The images are of model decelerating,
relativistic jets (the approaching jet is on the right). Vectors representing
the degree of polarization and the apparent magnetic field direction are
superposed on a grey-scale of total intensity. The location of the field
transition is marked. (a) Perpendicular-field spine and longitudinal-field shear
layer. (b) Field sheets wrapped around the jet axis, so that the
toroidal and longitudinal components are equal.}
\label{fig:fieldtrans}
\end{figure}

The other two field geometries sketched in Fig.~\ref{fig:fieldconfigs} can be
distinguished by the relative positions of the transition between longitudinal
and transverse apparent field on the jet axis in the main and counter-jets,
provided that the flow is relativistic and decelerating.  In the case of a
perpendicular-field spine surrounded by a longitudinal-field shear layer, the
apparent field transition always occurs further from the nucleus in the
counter-jet and may not be seen in the main jet.  For field sheets wrapped
around the jet axis, giving longitudinal and toroidal components only, the
converse is true.  This is illustrated in Fig.~\ref{fig:fieldtrans}.  It has
been known for some time that the brighter jet in an FR\,I source tends to show
a transition from longitudinal apparent field close to the nucleus to transverse
apparent field further out (Bridle 1984). Our deeper observations indicate that
the counter-jets either show transverse apparent fields on-axis over their
entire lengths, or have a transition much closer to the nucleus (Laing \& Bridle
2002; Canvin \& Laing 2004; Canvin et al.\ 2005; see also Hardcastle et
al. 1997).  We conclude that a first approximation to the field configuration in
FR\,I jets is a mixture of toroidal and longitudinal components.

\begin{figure}
\resizebox{\hsize}{!}  {\includegraphics[]{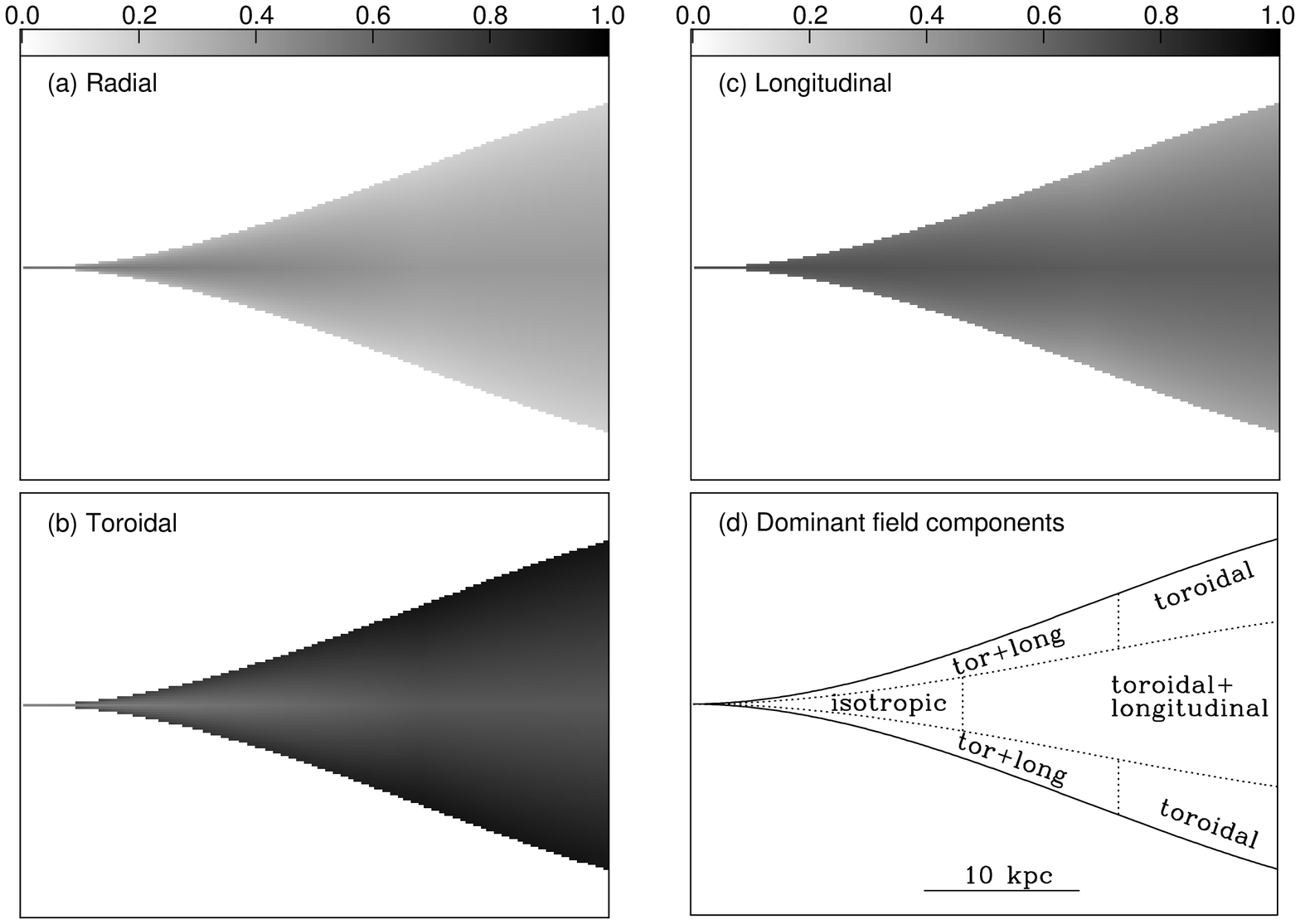}}
\caption{Panels (a) -- (c): grey-scales of the fractional magnetic-field
  components in NGC\,315 from Canvin et al. (2005). (a) Radial; (b) toroidal;
  (c) longitudinal. (d) Sketch of the dominant field components at different
  locations in the jet, as deduced from the model fits.}
\label{fig:bgrey}
\end{figure}

Detailed modelling (in which the ratios between the field components are allowed
to vary as functions of position) confirms this general conclusion, but adds
detail. Our conclusions for the four sources we have studied in detail are as
follows:
\begin{itemize}
\item Fields on kpc scales in FR\,I radio jets are not vector-ordered
  helices. Nor should they be: for a conical jet with velocity $\beta c$ and
  bulk Lorentz factor $\Gamma$, the fluxes for the longitudinal and
  transverse field components are $\propto r^{-2}$ and $\propto (\Gamma\beta
  r)^{-1}$, respectively, where $r$ is the distance from the nucleus. The
  longitudinal flux must be very small at large distances from the nucleus,
  otherwise the energy density in the magnetic field close to the accretion disk
  or black hole would be too high (Begelman, Blandford \& Rees 1984).
\item The field is primarily toroidal and longitudinal, with a smaller radial
  component in some objects.
\item The toroidal component could be ordered, provided that the longitudinal
  component has many reversals.
\item The longitudinal/toroidal field ratio decreases with distance from the
  nucleus, qualitatively as expected in an expanding, decelerating flow.
\item The evolution of the field-component ratios is not, however, consistent with flux
  freezing in a laminar velocity field of the type we infer, even if we include
  the effects of velocity shear (Laing \& Bridle 2004).
\end{itemize}
The field-component structure deduced from our model of NGC\,315 is shown in
Fig.~\ref{fig:bgrey}. 

\section{Further work}

In principle, our modelling technique should work for jets in powerful (FR\,II)
sources on kpc scales and for pc-scale jets. There are three main reasons why
this is difficult:
\begin{itemize}
\item counter-jets are faint and in some cases the jet/counter-jet ratios are
  very large (presumably the flow is faster);
\item the jets are narrow and
\item more powerful jets show significant sub-structure, over which we need to
  average in order to make a robust comparison of the main and counter-jets.
\end{itemize}
The only FR\,II jets which have been well enough resolved to give any idea of
their 3D field structures are those in 3C\,353 (Swain, Bridle \& Baum
1998). There, a roughly equal mix of longitudinal and toroidal field also gives
the best fit to the polarization data, although the jet velocity is not well
constrained.  A slightly different approach is feasible in micro-quasars, where
components from a single ejection event can often be identified on both sides of
the nucleus. Our technique could then be used to compare the brightness and
polarization of the jet and counter-jet components directly, with the added
advantage of known proper motions.

For these new applications, we will need sensitivities of 0.1 -- 1\,$\mu$Jy
rms and $>10^5$:1 dynamic range at frequencies chosen to be able to resolve and
remove Faraday rotation. The resolutions required range from 0.1\,arcsec for the
nearest FR\,II sources to $<0.1$\,milliarcsec for parsec-scale jets. We look
forward to EVLA, eMERLIN and broad-band VLBI.

\acknowledgements
 The National Radio Astronomy
Observatory is a facility of the National Science Foundation operated
under cooperative agreement by Associated Universities, Inc.

\end{document}